\documentclass[twocolumn]{svjour3}          
\usepackage{graphicx}
\usepackage{epstopdf}
\usepackage{lineno}
\begin{document}

\title{\boldmath Measurement of proton quenching in a LAB based liquid scintillator}

\author{Masheng Yang \and Zeyuan Yu \and Jun Cao \and Xilei Sun \and Boxiang Yu \and Guangpeng An}

\institute{
Zeyuan Yu \at
\email{yuzy@ihep.ac.cn}
\and
Masheng Yang \and Zeyuan Yu \and Jun Cao \and Xilei Sun \and Boxiang Yu \and Guangpeng An
   \at Institute of High Energy Physics, Beijing 100049, China
    \and
Masheng Yang \and Jun Cao
    \at University of Chinese Academy of Sciences, Beijing 100049, China
}
\date{\today}
\maketitle
\begin{abstract}
  Linear alkylbenzene has been recently used as the solvent of liquid scintillator by several neutrino experiments. The energy quenching effect of a linear alkylbenzene based liquid scintillator is studied in this paper with a 14 MeV D-T compact neutron generator, to improve the energy non-linearity modelling of this kind of detectors. The recoiled proton in the liquid scintillator has a kinetic energy ranging from 0.5 MeV to 13 MeV. The data is used to extract the parameters of the Birk's law, an empirical model to describe the energy quenching effect of the liquid scintillator.

\end{abstract}

\section{Introduction}
\label{sec:intro}
The linear alkylbenzene (LAB) based liquid scintillator (LS) has been widely used in current and future neutrino experiments, for example, Daya Bay~\cite{DYB-Detector}, RENO~\cite{RENO}, JUNO~\cite{JUNOYellowBook} and SNO+~\cite{SNO}. Energy response of the {LS} detector is not linear, partly due to the energy quenching effect of LS. The light output of the LS is related to the energy deposit density $dE/dr$. An empirical model, proposed by Birk~\cite{birksbook} and generalized by Chou~\cite{chou1952the}, is commonly used to describe the process, as shown in Eq.~\ref{formula1},
\begin{center}
\begin{equation}
\label{formula1}
  \frac{dL}{dr}=S\frac{\frac{dE}{dr}}{1+K_{B}\frac{dE}{dr}+K_{C}(\frac{dE}{dr})^2}\,,
\end{equation}
\end{center}
where $dL/dr$ is the light output, $S$ is the scintillation efficiency, $dE/dr$ is the energy deposit density,
 $k_B$ is Birks' constant, and $k_C$ is the second order parameter. Both $k_B$ and $k_C$ depend on the LS material.

The detector energy non-linearity is one of the key issues in a precise measurement of the reactor neutrino spectrum~\cite{DYBflux}. It will be even more important in the next generation neutrino experiment JUNO, which is designed to determine the neutrino mass hierarchy by measuring the small wiggles in the energy spectrum caused by neutrino oscillation. The expected energy spectrum without oscillation effect should be predicted to $\sim\,$1\% in shape uncertainty not to degrade the sensitivity to the mass hierarchy. Therefore, the detector energy non-linearity must be carefully studied.

Recently there have been several experiments reporting benchtop measurements\cite{Feihong}\cite{Berkely}\cite{BNL} and in-situ fitting to the quenching effects~\cite{SoerenNeutrino2014}\cite{DCMeasurement}. The Birks' constant $k_B$ seems to be significantly different between protons and electrons. This phenomena hasn't been well explained. A comprehensive measurements with different particles will give a whole picture of the energy quenching effects of the LS.

In this article, we present the measurement to proton quenching effects with a collimated 14 MeV Deuterium-Tritium neutron generator and a benchtop experimental setup. Neutron from the generator interacts with the proton in the LS mainly via the elastic scattering. Special care was taken of the energy scale calibration and uncertainty estimation.

\section{Experimental setup}
\label{sec:setup}
\subsection{Detector and electronics}

\begin{figure}
  \centering
  \includegraphics[width=8cm]{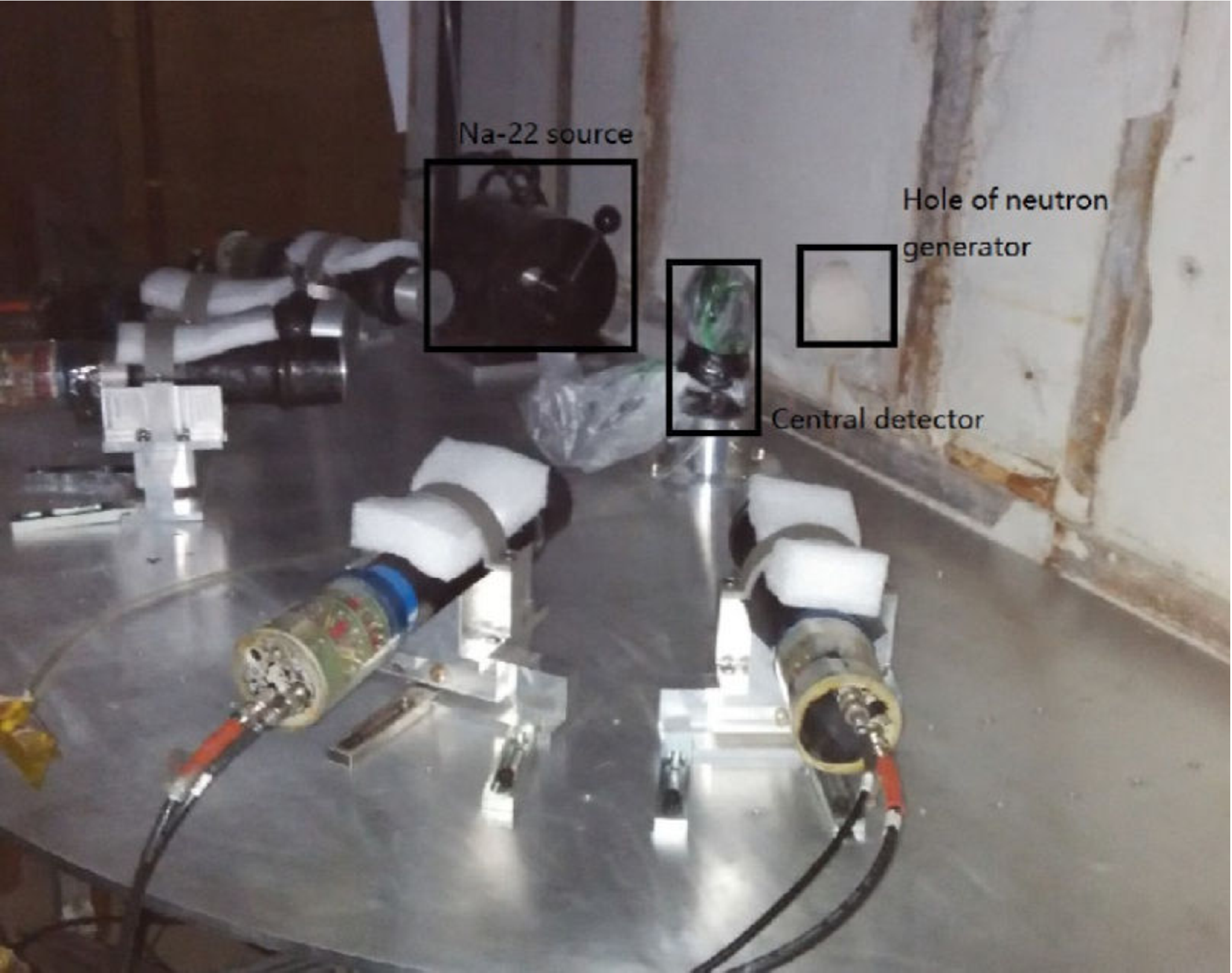}
  \caption{ Experimental setup. A Deuterium-Tritium 14 MeV neutron generator serves as the neutron source. An collimated and intensive $^{22}$Na source serves as the energy calibration source. The $\gamma$ direction is approximately perpendicular to the neutron beam. }
  \label{fig:setup}
\end{figure}

A 14 MeV Deuterium-Tritium compact neutron generator was used as the neutron source. It was surrounded by 1 meter thick concrete wall for protection. A hole of 8.5 cm in diameter served for the neutron collimation. Detectors were fixed on a semicircular table facing to the collimation hole. The experimental setup is show in Fig.~\ref{fig:setup}. Direction of the collimated $^{22}Na$ $\gamma$ was approximately perpendicular to the neutron beam, however the precise direction of neutron beam was not known.

The LS sample was contained in a cylindrical quartz vessel with a diameter and height of 5 cm. A XP2020 photomultiplier tube (PMT) was coupled to the quartz vessel for the scintillation photon detection. It was named as the central detector in this paper.

Five BC501A LS detectors were placed around the central detector and their distances to the central detector were about 50~cm. They were named as coincidence detectors. BC501A LS has good neutron and gamma discrimination ability to eliminate the backgrounds. Each coincidence detector was also coupled to a XP2020 PMT. All six detectors were tightly wrapped up with black adhesive tape to prevent light leakage, and firmly fixed on the table with brackets.

\begin{figure}

  \centering
  \includegraphics[width=8cm]{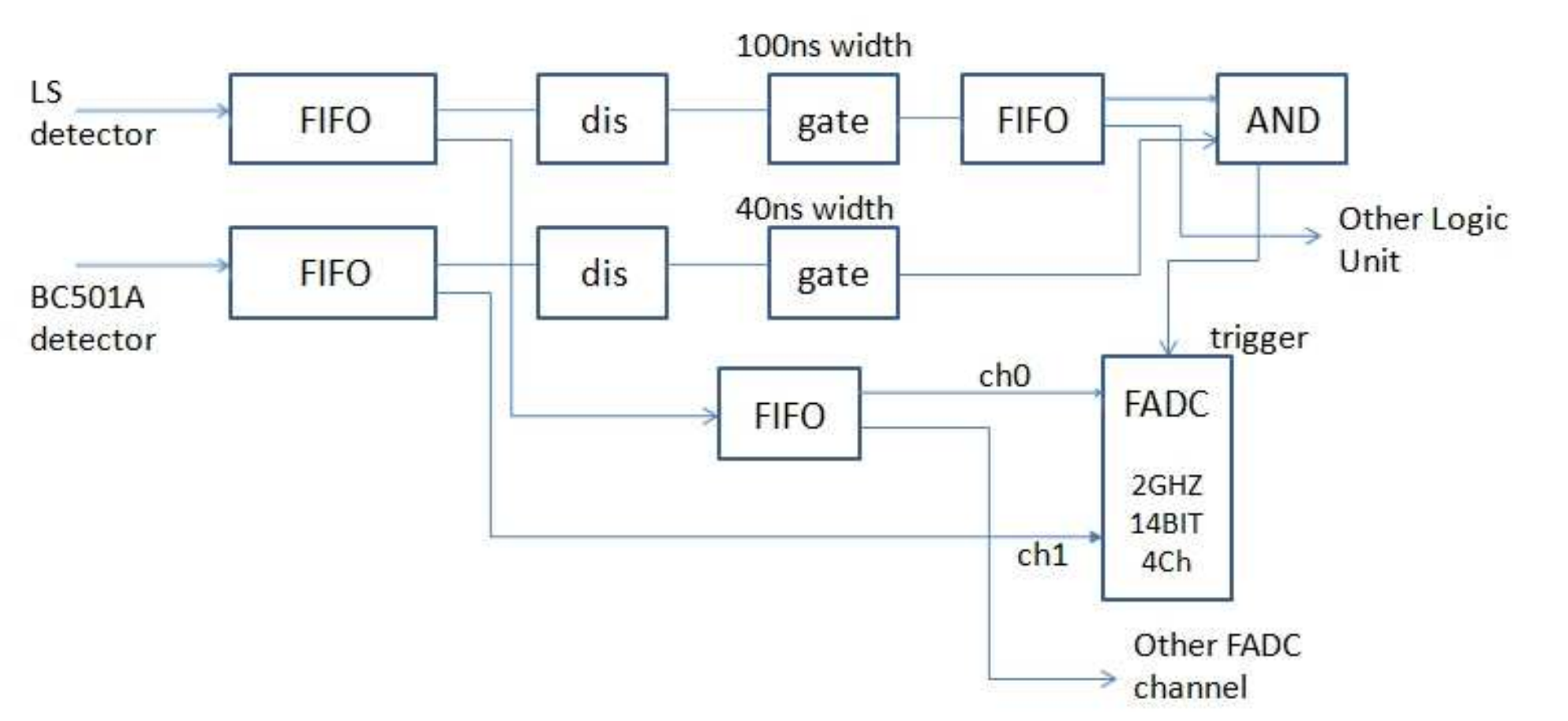}
  \caption{ Scheme of the electronics system.}
  \label{fig:electronics}
\end{figure}

An NIM-based electronics system was set up for the trigger and DAQ. Signals from the central LS detector and five coincidence detectors were sent to a Constant Fraction Discriminator (CFD) for the single channel trigger. If the central detector and one coincidence detector passed the threshold in a 500 ns time window, the event, with the data from six channels, was recorded by two FADCs boards (CAEN V1729 Digitizer, 4-channel, 12-bits resolution, 2 GHz sampling frequency). The scheme of the trigger and electronics is shown in Fig.~\ref{fig:electronics}.

\subsection{Data taking}
\label{Datataking}

The principle of the measurement was the elastic scattering between neutrons and protons in the LAB based LS.
The recoiled proton's energy was estimated via the scattering angle.
In order to cover the proton energy from 0.5 MeV to 13 MeV, five rounds of measurements were taken in 10 days.
Tab.\ref{angles} shows the scattering angles at which the coincidence detectors were placed in the 5 rounds of measurements.
"PMT1","PMT2",... "PMT5" are the names of the 5 coincidence detectors.
One of the five coincidence detectors(PMT2) was fixed at 40$^{\circ}$,
 used to monitor the possible gain shift of the central detector during the 5 measurements.
Finally, we managed to cover the scattering angles from 25$^{\circ}$ to 75$^{\circ}$ every $5^{\circ}$ .

\begin{table}
{
\begin{tabular}{|c|c|c|c|c|c|}
  \hline
         & PMT1 & PMT2 & PMT3 & PMT4 & PMT5 \\
  \hline
  round~1 & -60 & -40 & -15 & 40 & 60 \\
  \hline
  round~2 & -65 & -40 & -20 & 35 & 65 \\
  \hline
  round~3 & -55 & -40 & -25 & 45 & 70 \\
  \hline
  round~4 & 15 & -40 & 30 & 50 & 75 \\
  \hline
  round~5 & 25 & -40 & 55 & - & - \\
  \hline
\end{tabular}
  \caption{Scattering angles of the 5 coincidence detectors during the whole 5 rounds of measurements(unit: degree), The "$-$" means the coincidence detector was placed on the left side of the beam.\label{angles}}
}
\end{table}

The temperature in the laboratory was monitored and kept stable to $\pm$0.5~$^{\circ}$C during the whole data taking period. Therefore, the temperature impact to the LS quenching was negligible. Before each measurement, the LS sample in the central detector was bubbled with pure nitrogen for 1 hour to purge the oxygen, which could introduce another kind of quenching effect\cite{Xiaobo}.
\subsection{Calibration}
\label{Calibration}

In general, the quenching experiment compared the proton's visible energy and true energy. The true energy was estimated with the scattering angle, and the visible energy was reconstructed with an electron energy scale.

The energy scale was calibrated with Compton electrons from the $^{22}Na$ $\gamma$ source, which provided intensive and collimated gamma rays, shooting the LS sample from the direction perpendicular to the neutron beam. The $\gamma$ generated an electron in the central detector via the Compton scattering. The coincidence detectors can detect the scattered $\gamma$, and the energy of Compton electron was determined with the scattering angle. Compared to the traditional calibration method using the backward scattering Compton edge which has an about 5\% uncertainty, this energy scale calibration method could reach better than 1\% precision.

Since the proton (electron) true energy was estimated with the neutron ($\gamma$) scattering angle, the angle should be measured precisely. The angle between the collimated $\gamma$ beam, the central detector and coincidence detector was carefully measured with a laser gradient with a precision of 0.3$^{\circ}$, and cross checked with a digital protractor of precision 0.2$^{\circ}$. The two measurements were consistent within 0.3$^{\circ}$, which was quoted as uncertainty of the scattering angle.

However the angle between neutron beam and $\gamma$ beam couldn't be measured precisely and it was approximate 90$^{\circ}$. This introduced a correlated uncertainty to proton energy calculation at different angles. Since the coincidence detectors can be placed both at the left and right of the neutron beam, measurements at five angles were done at both sides. In this way, the angle uncertainty of neutron beam could be cancelled, because this uncertainty had oppositive influences to the right and left scattering.

\section{Data analysis}
\subsection{Event selection}
\label{sec:event}

The coincidence detectors with BC501A liquid scintillator~\cite{lsdis}) were used to detect scattered neutrons. To reject coincidence backgrounds from the ambient $\gamma$s or $\gamma$s generated by neutron capture on nuclei, pulse shape discrimination (PSD) was used, since more slow scintillation photons were produced by proton (from neutron scattering) than by gamma and electron in the BC501A.

\begin{figure}
  \centering
  \includegraphics[width=8cm]{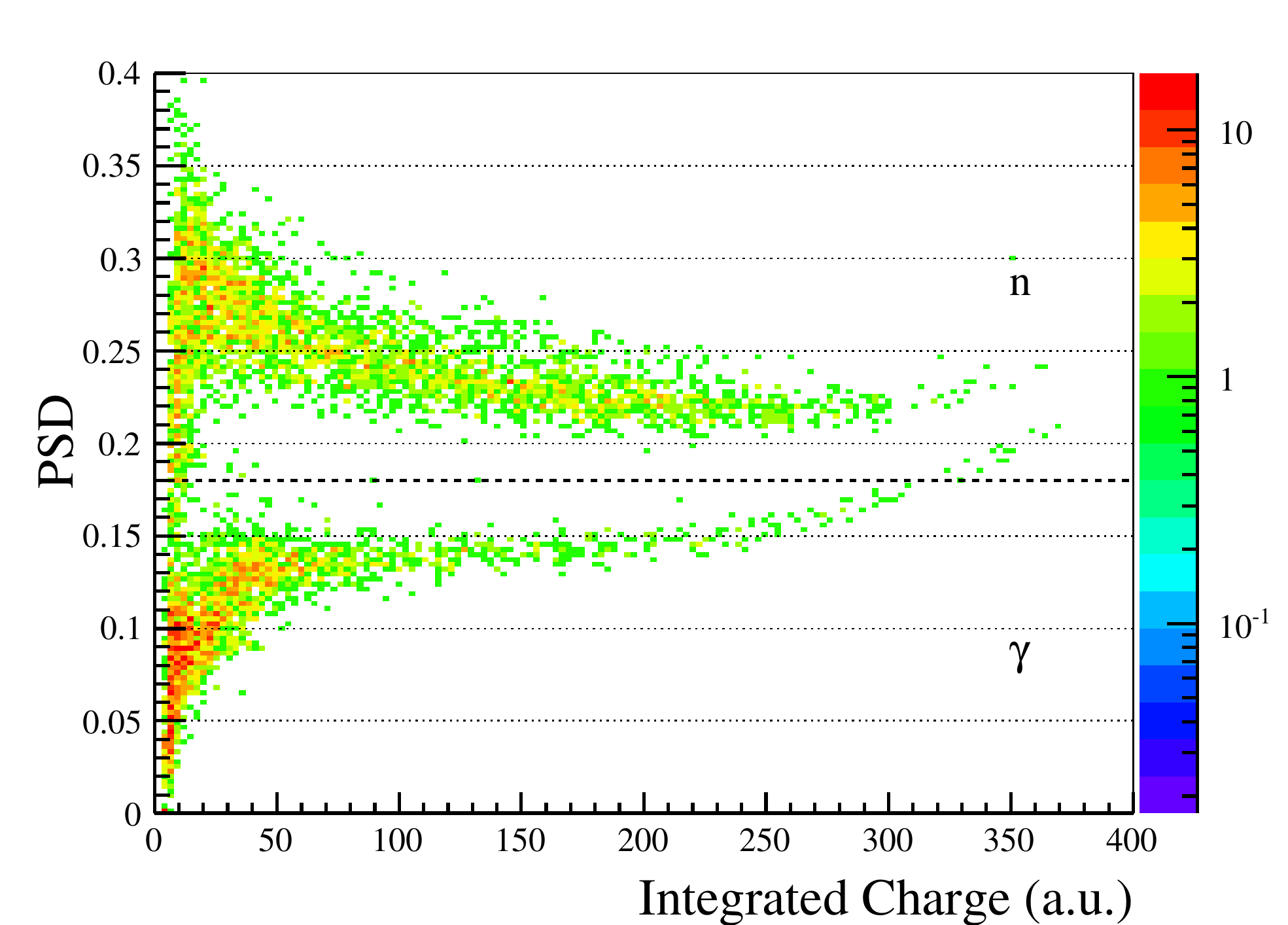}
  \caption{ PSD for $n/\gamma$ discrimination in the BC501A detectors. Events in the upper part are neutrons and those in the lower part is $\gamma$.}
  \label{fig:Ipairs}
\end{figure}

A pulse shape discriminator was defined as $\rm{PSD}=I_{\rm delay}/I_{\rm total}$, where $I_{\rm delay}$ was the integrated charge in the time window [T$_{0}$+35 ns, T$_{0}$+150 ns], $I_{\rm total}$ was that in [T$_{0}$, T$_{0}$+150 ns], with T$_{0}$ the trigger time. The discriminator PSD was shown in Fig.~\ref{fig:Ipairs} as a function of the integrated charge. Events in the upper part were identified as neutron, and events in the lower part were identified as $\gamma$s.

\begin{figure}
  \centering
  \includegraphics[width=8cm]{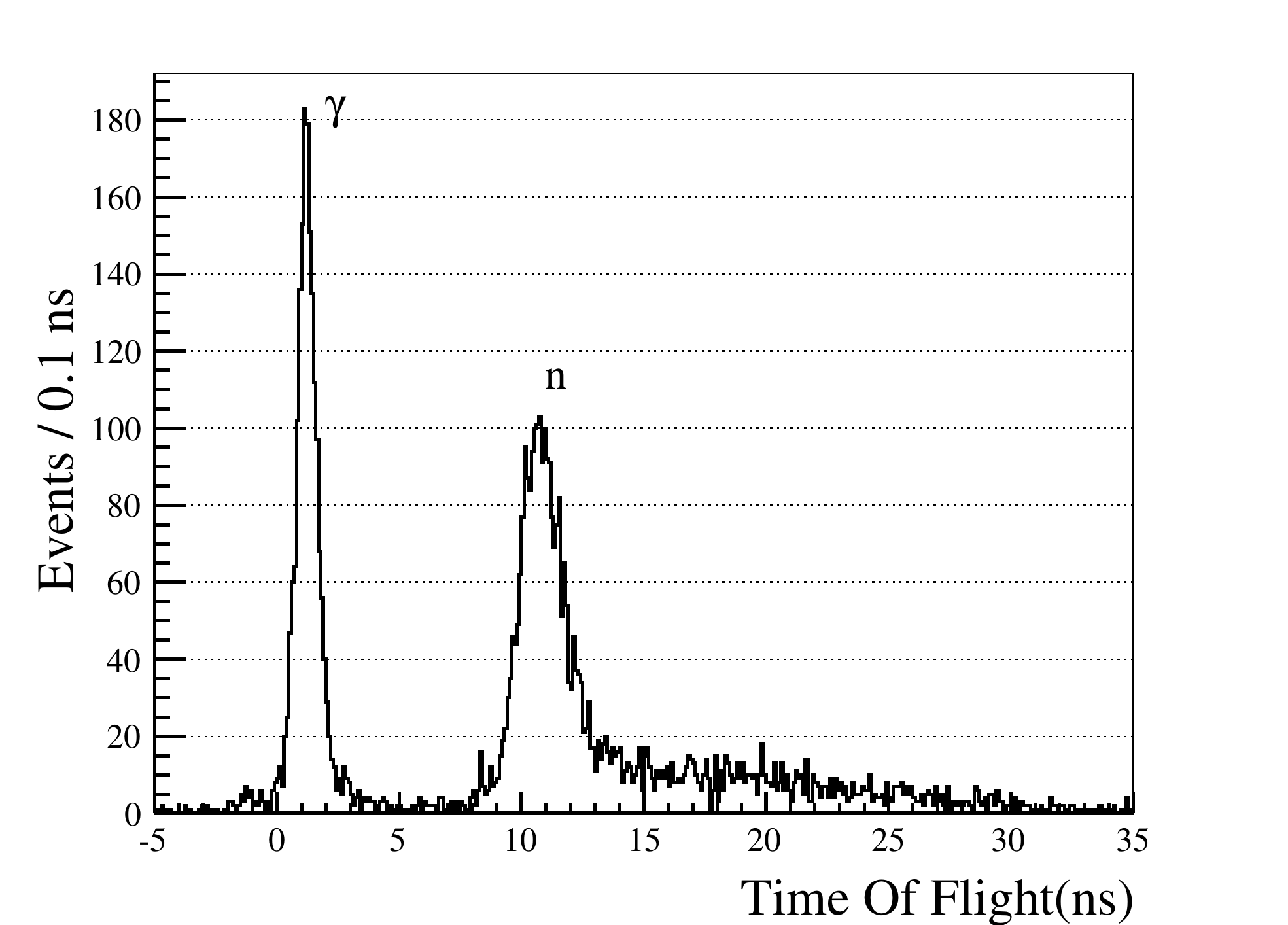}
  \caption{ Distribution of the time of flight. The left peak is from $\gamma$ Compton scattering and the right one is from elastic scattering neutrons.}
  \label{fig:TOF}
\end{figure}

Comparing to $\gamma$, elastic scattering neutron at a certain angle had a certain delay in the time of flight (TOF). As shown in Fig.~\ref{fig:TOF}, the left peak was $\gamma$ and the right one was neutron. The tail larger than 15 ns was mainly due to the inelastic scatted neutrons on $^{12}$C, and multiple scattering neutrons which were scattered more than once. In both cases the energy of outgoing neutron was lower than elastic scattering neutron and it spent longer time during the flight from the central detector to the coincidence detector.

\begin{figure}
  \centering
  \includegraphics[width=8cm]{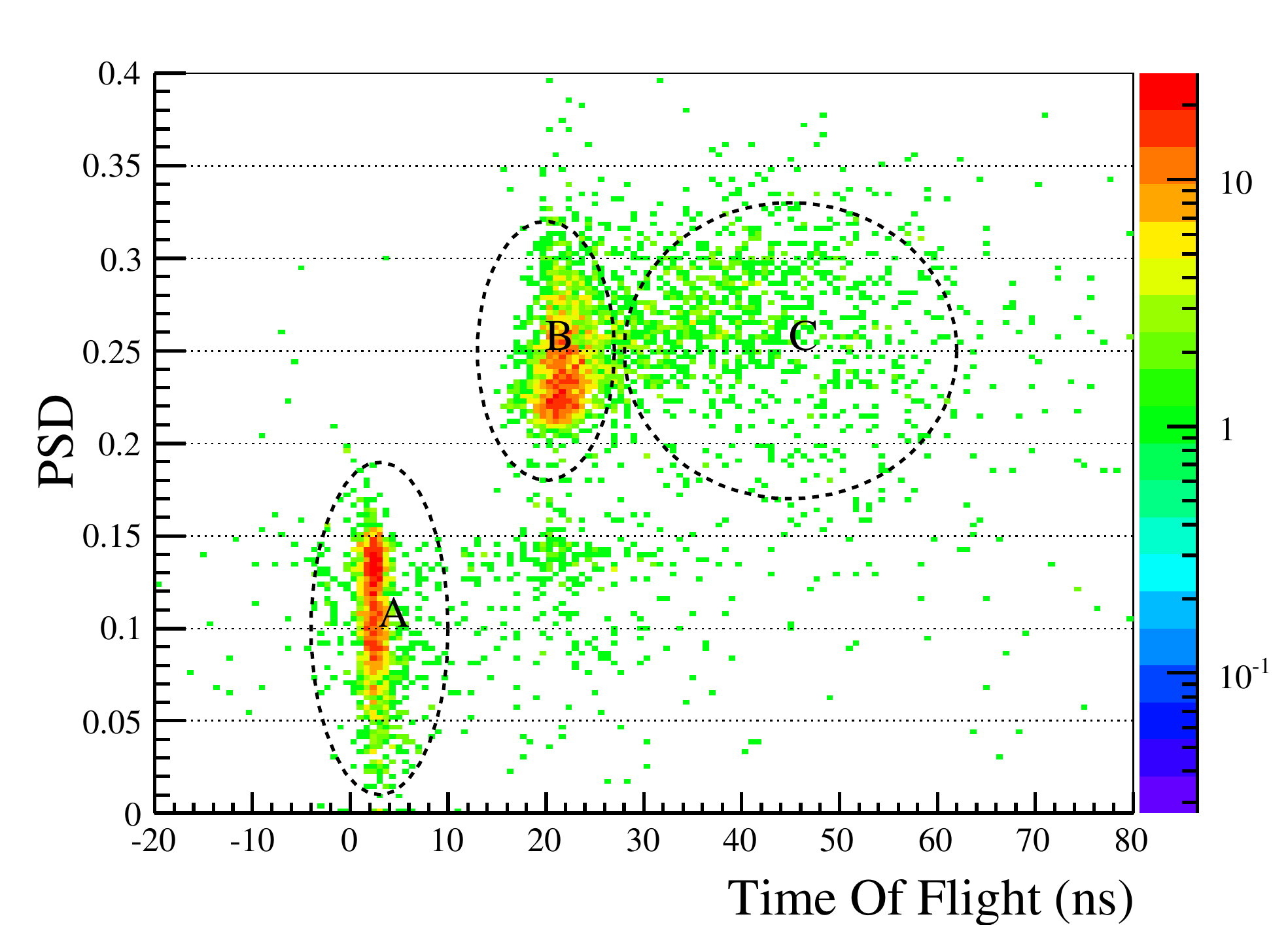}
  \caption{ Distribution of the TOF versus PSD. Events in Region B are from n-p elastic scattering, and those in region A are from $\gamma$ Compton scattering.}
  \label{fig:psd-tof}
\end{figure}

The final selection was based on the combination of PSD and TOF, as shown in Fig.~\ref{fig:psd-tof}.
Events in region B were the n-p elastic scattering events, and those in region A were the gamma Compton scattering events. Events in region C were mainly the multiple scattering and inelastic scattering neutrons.

The major background, accidentals, which came from the accidental coincidence of events between the central detector and the coincidence detectors, were numerically calculated with the single event rates and the trigger window length, and found to less than 0.1\% comparing to the 0.5 Hz neutron scattering rate. Thus the background was neglected in the analysis.

\subsection{Energy reconstruction}

In general, the quenching experiment compared the particle's visible energy and true energy.

The true energy was determined with the scattering angles. As discussed in Sec.~\ref{Calibration}, uncertainty of the angle was 0.3$^{\circ}$. The measurements at both left and right sides cancelled the uncertainty of neutron beam direction.

The visible energy, which was defined as the energy converted to scintillation photons, was reconstruction following two steps:

1) PMT charge integration. The total PMT charge was proportional to the particle's visible energy, and it was calculated by the waveform integration method. The integration started from 30 ns before the trigger and ended when the waveform returned to the baseline. The XP2020 PMT and its well designed base gave us a beautiful waveform, without overshoot, reflections and ringing etc. According to waveform simulation and reconstruction method described in~\cite{Zeyuan}, the charge integration non-linearity was determined to be less than 0.2\%, which was negligible.

2) Divide the PMT charge with energy scale, which was determined by means of the 0.321 MeV electron from $^{22}$Na's 0.511 MeV $\gamma$. Although relatively small, the LS detector also has non-linearity response to electrons. The 0.321 MeV electron was chosen due to that at this energy, simulation shows the quenching effect is less than 3\% and the Cherenkov light contribution is less than 0.2\%.

\begin{figure}
  \centering
  \includegraphics[width=8cm]{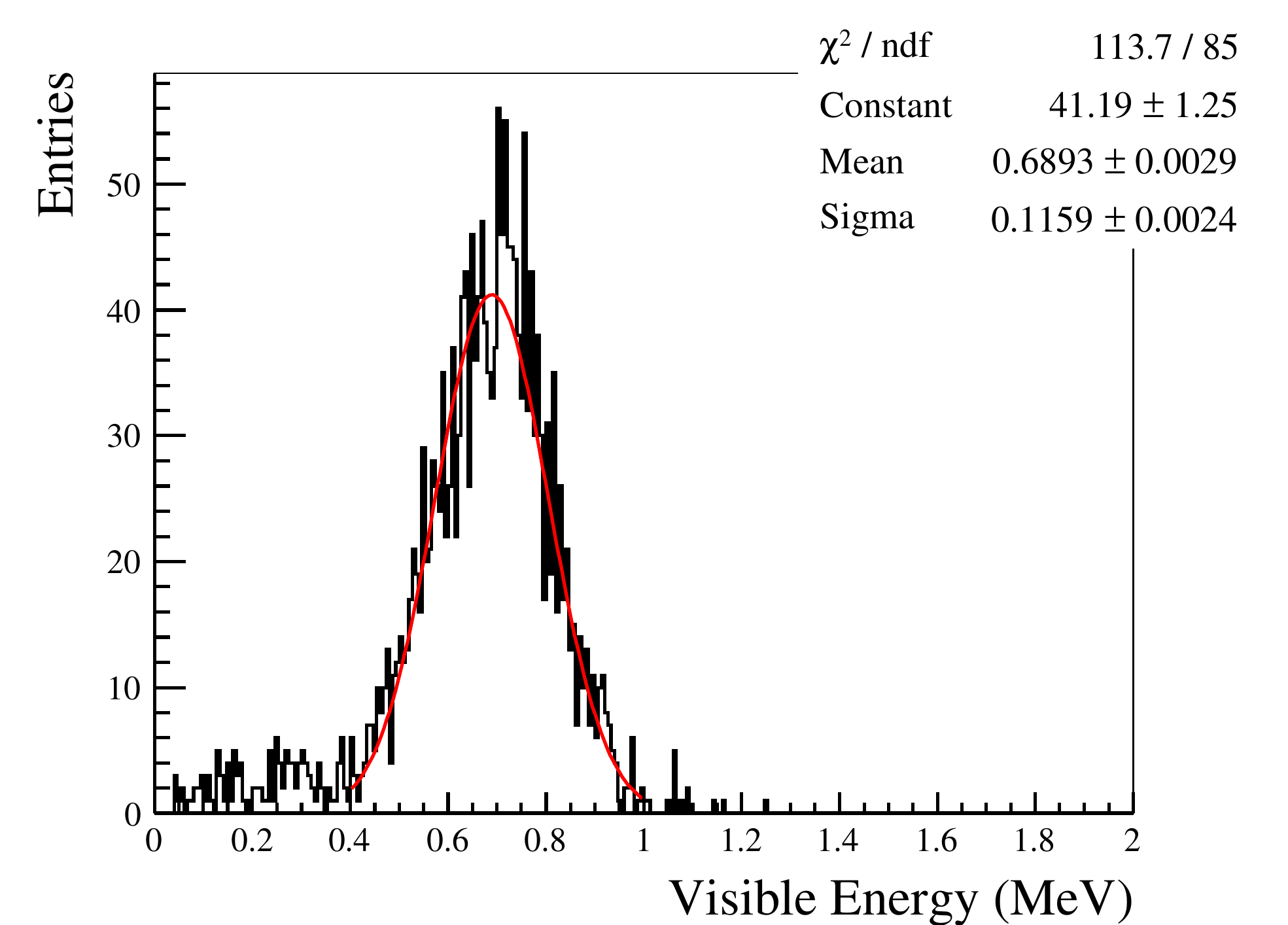}
  \caption{ An example of visible energy spectrum of the selected recoil proton events in the central detector.}
  \label{fig:Proton-spec}
\end{figure}

\begin{figure}
  \centering
  \includegraphics[width=8cm]{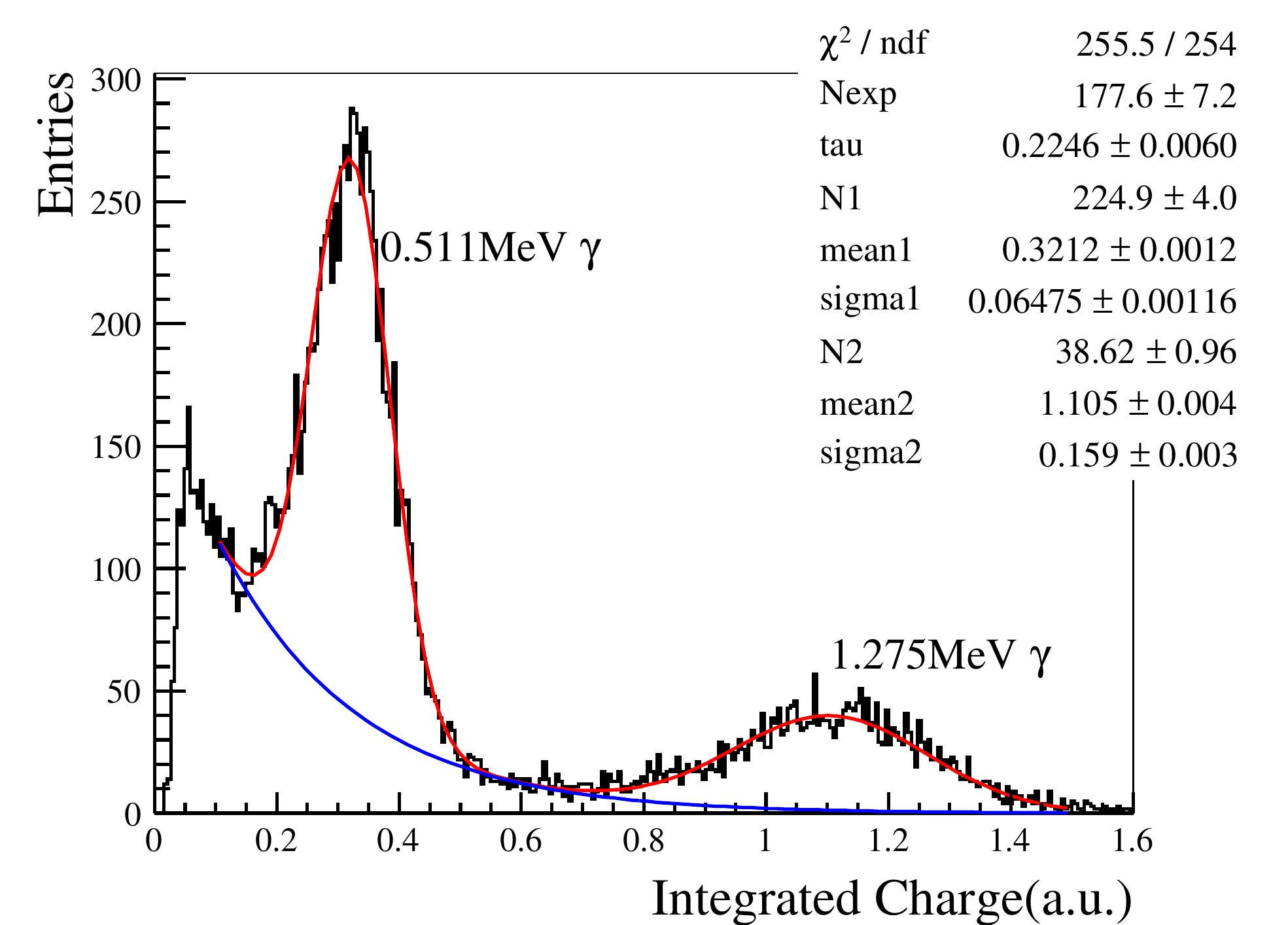}
  \caption{Energy spectrum of the Compton electrons, fitted with exponential background (blue) plus two Gaussians. The left peak was electrons from the 0.511 MeV $\gamma$ and the right one was from 1.27 MeV $\gamma$.}
  \label{fig:ElecCalib}
\end{figure}

Fig.~\ref{fig:Proton-spec} shows an example of the visible energy spectrum of the selected proton events in the central detector. A Gaussian function was used to extract the peak value. Fig.~\ref{fig:ElecCalib} shows the energy spectrum of Compton electrons, fitted with exponential background (blue) plus two Gaussian functions. The right peak came from the Compton scattering of 1.27 MeV $\gamma$ while the left peak was from the scattering of 0.511 MeV $\gamma$.

\section{Fitting with Birks' law}
\label{sec:rst}

\subsection{Construction of the $\chi^2$ function}

Since several systematic uncertainties were correlated, such as neutron beam angle uncertainty, electron energy scale uncertainty, a least square method with pull terms was used to the parameter $k_B$ and $k_C$. The $\chi^2$ function was constructed as

\begin{eqnarray}\label{qf}
  \chi^{2}=\sum_{i=1}^{N} \frac{ (Q_i-f_e\frac{Q_e}{E_e} f_{q}(k_B,k_C,E_{i}) E_i)^{2} }{\sigma_{Q_{i}}^{2}}+
  \\
  \frac{\delta^2}{\sigma^2_{\delta}}+\frac{(f_e-f_e^{\rm MC})^2}{\sigma^2_{e}},
\end{eqnarray}

where $Q_i$ was the measured proton charge at different angles labelled by $i$, E$_i$ was the calculated kinetic energy of the recoil proton. Q$_e$ and E$_e$ were the measured charge and the calculated true energy of the Compton electron, respectively. $f_e$ was a pull term to take into account the uncertainty of the electron energy scale calibration, and $f_e^{\rm MC}$ was the predicted non-linearity from Monte Carlo.

$E_i$ was calculated with
\begin{center}
\begin{equation}\label{penergy}
    E_{i}=E_{n} sin^{2}(\theta_i\pm\delta)\,
\end{equation}
\end{center}

where $\delta$ was a pull term taking into account uncertainty of neutron beam direction. The $\pm$ meant the coincidence detector was placed on the right or left side of the beam. The uncertainty of $\delta$ was taken as 0.5 degree according to the construction manual of the concrete wall.

The quenching function $f_{q}(k_B,k_C,E_{i})$ was integrated with the Birks' law

\begin{equation}\label{birkslaw}
    f_{q}(k_B,k_C,E_{i})=\int_{0}^{E_i} \frac{dE}{1+k_B(\frac{dE}{dr})+k_C(\frac{dE}{dr})}.
\end{equation}

The integration was done numerically. In each step, step length was chosen to be 1 $\mu m$ and the energy loss $dE$ in this step was calculated with $\frac{dE}{dr}$. $\frac{dE}{dr}$ was the proton energy deposit density estimated with the frequently used software SRIM~\cite{SRIM}. After this step, dE was subtracted from the proton energy, and the integration was finished when the proton energy was equalling or less than zero.

\subsection{Systematics uncertainty}

The systematics uncertainty of $\sigma_{Q_{i}}$ had the following four contributions:

1) Event selection and spectrum fitting. The TOF cut was varied and the fitting function was changed to the CrystalBall function. The fitting results were consistent 0.5\% which was taken as uncertainty.

2) Detector system stability. The whole data taking had five rounds and the fixed 40$^o$ coincidence detector was used to correct the stability. Uncertainty of the correction came from the statistic fluctuations of the normalization points which was about 0.2\%. This uncertainty was correlated within one round and uncorrelated between different rounds. To simplify the fitting, the uncertainty was treated as fully uncorrelated but increased to 0.5\%.

3) The possible non-linearity effect in the PMT and the FADC system. The PMT high voltage was carefully adjusted to ensure that PMT was working in the linear region. The FADC linearity was tested with a pulse generator, and no obvious non-linearity was found. Regarding to the data sheets of PMT and FADC, and the waveform analysis methods, a conservative 0.5\% uncorrelated uncertainty was set to each data point.

4) Scattering angles. This uncertainty belonged to the proton true energy $E_i$, but since it was uncorrelated among different angles, it was absorbed to $\sigma_{Q_{i}}$. The angle uncertainty was 0.3$^o$ and was propagated to $\sigma_{Q_{i}}$ at each data point.

The $\sigma_e$ which was uncertainty of electron energy scale had two components:

1) Electron quenching. MC gives the electron quenching factor was about 3\% at $k_B$=7.0$\times$10$^{-3}$~g/cm$^2$/MeV. During the fitting, when scanning the $k_B$ and $k_C$ parameter space, the factor was also changed. Uncertainty of the factor was estimated to be less than 0.3\%, mainly raised from the different electron energy deposit density modeling.

2) Electron spectrum fitting. Varying the fitting range and fitting function gave 0.4\% changes in the 0.321 MeV electron peak fitting. The 0.4\% was taken as uncertainty.

The both components are correlated among all proton data points and they were combined to get the $f_e$

\subsection{Fitting results}

\begin{figure}
  \centering
  \includegraphics[width=8cm]{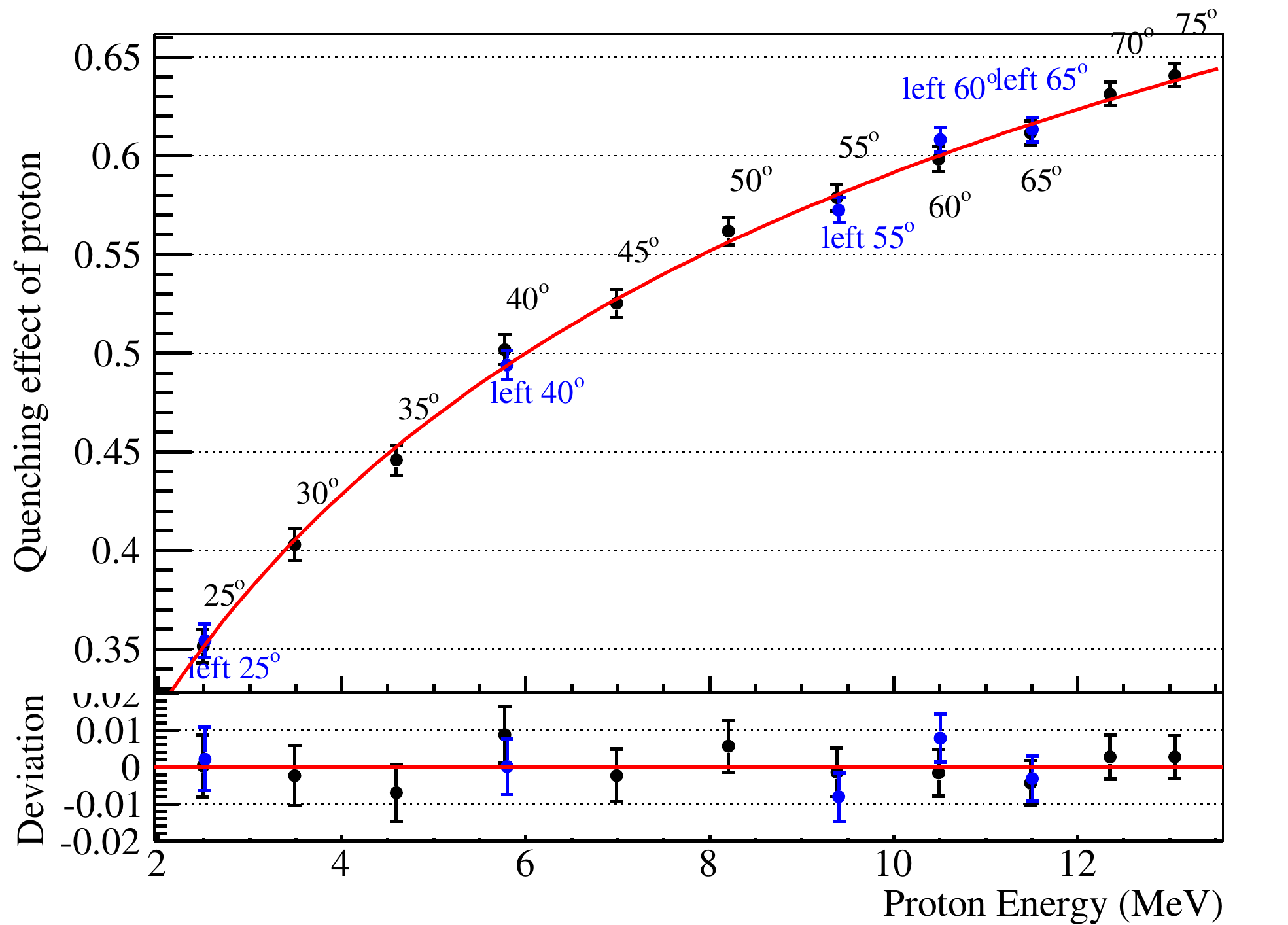}
  \caption{Proton's quenching best fit curve and the $k_B$ equals to (7.50$\pm$ 0.19)$\times$10$^{-3}$~g/cm$^2$/MeV.}
  \label{protonFinalresult}
\end{figure}

The ROOT TMinuit package was utilized to minimize the $\chi^2$ function, the best fit Birks' constant was $k_B$=(7.50$\pm$ 0.19)$\times$10$^{-3}$~g/cm$^2$/MeV and
the second order parameter  $k_C$=(2.03$\pm$1.0)$\times$10$^{-6}$~g$^2$/cm$^4$/MeV$^2$.
The fitting quality was $\chi^{2}_{\rm min}$/NDF=7.30/13, and the best fit neutron beam angle deviation is $\delta$=0.05$^{o}$. The best fit quenching curve and data points are shown in Fig.~\ref{protonFinalresult}.

It should be noted that the proton quenching was less sensitive to the $k_C$ term, thus the $k_C$ had much larger uncertainty than $k_B$. For example, with the best fit Birk's constants, the $k_C$ term only contributed about 4\% to the quenching if the proton energy was 10 MeV. To determine $k_C$ precisely, the $\alpha$ particle were required since its energy deposit density was much large.

The last systematic uncertainty source was the modeling of proton energy deposit in LS. The SRIM software was the most frequently used one considering the Ziegler, Biersack and Littmark stopping model in matter. In our study, Geant4~\cite{Geant4} was also used to get the dE/dr and served as a crosscheck. The fitting results were consistent to less than 1\%.

\section{Conclusion}
\label{sec:conclusion}

The proton's response in a LAB-based liquid scintillator was measured with a 14 MeV monogenetic neutron generator. With coincidence detectors at different angles from 20$^{o}$ to 75$^{o}$, the measurement covered proton energies from 0.5 MeV to 13 MeV. Systematic uncertainty was carefully studied and a special care was taken for the neutron beam direction uncertainty.
The current benchtop measurement and in-situ calibration results indicate that the $k_B$ are different between proton and electron. However, the phenomena hasn't been well explained and need further study. This measurement will contribute to the thorough understanding of LS non-linearity.

\begin{acknowledgements}
This work was supported by National Natural Science Foundation of China
(Grant No.11390381 and 11225525).
\end{acknowledgements}

\end{document}